\documentclass[onecolumn]{aa} 
\usepackage{natbib}

\usepackage{float}
\usepackage{xcolor}
\usepackage{mathptmx}
\usepackage{amsmath}
\usepackage{graphicx}
\usepackage{array}
\usepackage{multirow}
\usepackage[unicode=true,
 bookmarks=true,bookmarksnumbered=false,bookmarksopen=false,
 breaklinks=false,pdfborder={0 0 0},backref=false,colorlinks=true]
 {hyperref}

\usepackage{breakurl}

\makeatletter

\begin{document}

\title{Characterizing spectral continuity in SDSS u'g'r'i'z' asteroid photometry}
\subtitle{}

\author{P. H. Hasselmann\inst{1,2}
             \and
             M. Fulchignoni\inst{2}
            \and
            J. M. Carvano\inst{1}
            \and
            D. Lazzaro\inst{1}
            \and M. A. Barucci\inst{2}
            }

\institute{Observat\'orio Nacional (COAA), Rua General Jos\'e
Cristino 77, 20921-400, S\~ao Crist\'ov\~ao, Rio de Janeiro RJ, Brazil.
\and
LESIA, Observatoire de Paris, CNRS, UPMC, Universitaire Paris-Diderot, 5 place Jules Janssen, 92195 Meudon, France.}

 \authorrunning{ Hasselmann  et. al}

   \date{}


  \abstract
   {
   The 4th release of the SDSS Moving Object Catalog (SDSSMOC) is presently the largest photometric dataset of asteroids.
   Up to this point, the release of large asteroid datasets has always been followed by a redefinition of asteroid taxonomy. 
   In the years that followed the release of the first SDSSMOC, several classification schemes using its data were proposed, all using the taxonomic classes from previous taxonomies.  However, no successful attempt has been made to derive a new taxonomic system directly from the SDSS dataset.
   }
   {
    The scope of the work is to propose a different interpretation scheme for 
    gauging u'g'r'i'z' asteroid observations based on the continuity of spectral features. 
    The scheme is integrated into previous taxonomic labeling, but is not dependent on them.
   }
   {
   We analyzed the behavior of asteroid sampling through principal components analysis 
   to understand the role of uncertainties in the SDSSMOC.
   We identified that asteroids in this space follow two separate linear trends using reflectances in the visible, which is characteristic of their spectrophotometric features.
   }
   {
    Introducing taxonomic classes, we are able to interpret both trends as representative of featured and featureless spectra. 
    The evolution within the trend is connected mainly to the band depth for featured asteroids and to the spectral slope for featureless ones. 
    We defined a different taxonomic system that allowed us to
only classify asteroids by two labels.
   }
   {
    We have classified 69\% of all SDSSMOC sample, which is a robustness higher than reached by previous SDSS classifications. 
    Furthermore, as an example, we present the behavior of asteroid (5129) Groom, whose taxonomic labeling changes according to one of the trends owing to phase reddening.
    Now, such behavior can be characterized by the variation of one single parameter, its position in the trend.
   }


   \maketitle
%

\section{Introduction}

Characterization of the compositional versus dynamical
distribution of the asteroid population has relied on large datasets.
In particular, the definition of asteroid classes, or taxonomies, has
been historically linked to the development of new sets of asteroid
observations. The first taxonomy, called CMZ
\citep{1975Icar...25..104C}, began with the TRIAD catalog
(\emph{Tucson Revised Index of Asteroid Data},
\citealp{1978Icar...35..313B}). This dataset was composed of
radiometric, photometric, and polarimetric measurements of 110
asteroids obtained from 1972 to 1978
\citep{TRIADRAD}. The derived taxonomy divided
the asteroids among S (\emph{``Stone}''), C (\emph{``Carbonaceous}''),
and U (``\emph{Unclassified}''), plus a single type representing
Vesta alone.

The second generation of taxonomies, in the mid-1980s, was boosted by the
\emph{Eight-Color Asteroid Survey} (ECAS,
\citealt{1985Icar...61..355Z}) and gave birth to three taxonomic
schemes \citet{1984PhDT.........3T}, \citet{1987Icar...72..304B}, and \citet{1989AJ.....97..580T}. 
The so-called Tholen taxonomy
\citep{1989aste.conf..298T} used principal component analysis and
minimal tree techniques to divide the asteroids into 14 classes
\textemdash S, C, G, B, F, A, D, T, E, M, P, R and V \textemdash
defined by their IRAS albedo and colors.

A third generation of taxonomies came with the advent of asteroid
spectroscopy. Using visible spectra from
the Small Main-belt Asteroid Spectroscopic Survey II (SMASS II) and a
methodology similar to Tholens' (plus visual inspection for spectral features),  \citet{2002Icar..158..146B} 
increased the number of classes to 26. 
Some years later, a near infrared (NIR) extension to this system,
based on data for 371 asteroids \citep{2008Icar..194..436D}, was
developed by \citet{2009Icar..202..160D}, decreasing the number of
classes to 23.

More recently, the release of the Sloan Digital Sky Survey (SDSS,
\citealp{2002AJ....123..485S}) and its subproduct, the Moving Objects
Catalog (SDSSMOC, \citealp{2001AJ....122.2749I,2002SPIE.4836...98I}),
allowed the development of several studies analyzing the distribution
and taxonomic classification of asteroids using the four colors of the survey.
This is the largest asteroid color catalog available with about four hundred thousand detections and reaching asteroids as small as 1km. Up to now, several authors have applied  a myriad of different approaches in order to use its data to classify asteroids according to previous taxonomies
\citep{2001AJ....122.2749I,2005Icar..173..132N,2006Icar..183..411R,2008Icar..198..138P,2010A&A...510A..43C}. 
Even though these SDSS taxonomic classifications  have  proven to be extremely useful in characterizing the distribution of spectral types in the main belt, they also end up hiding much of the spectral diversity that may be  present within each class (and between classes) in the SDSS dataset. Also, these methodologies are unable to detect any new spectral type that may exist in the sample.

An attempt to derive a new taxonomic system directly from the SDSS dataset was presented in a previous work  \citep{2013scpy.conf...48H}. In this paper, a modified version of the G-mode clustering algorithm was applied to the SDSSMOC.  The G-mode algorithm is an unsupervised classification method that depends on only one free parameter, the confidence level q1.  The final number of classes obtained varied with the adopted value of q1 from 1.0  to 3.0. However, subsequent  tests showed that the classes obtained in this way were of limited practical use, since either the classes obtained were too inclusive in terms of spectral shapes or the number of unclassified samples ended up being  too high.

The lack of success of this attempt is due to  two characteristics of the SDSS dataset. One is the relatively high uncertainties of the observations of the fainter objects, which dominate the sample. The other is the sheer number of the observations in the dataset. Their combined effect is to blur the distinction between the classes.


Therefore, our goal here is to propose a different approach to interpreting \emph{u'g'r'i'z'} asteroid photometry. We investigate the continuity of spectral features and propose a more suitable scheme based on this assumption and only relying on the information given by the dataset. To interpret the proposed scheme, we then integrate it with the traditional taxonomies. We regard the present work mainly as exploratory, where the techniques used here are applicable to other datasets, such as the ones that will be generated by the \emph{Gaia} mission  \citep{2012P&SS...73...86D} and the LSSST \citep{2014DPS....4621413J}. 

A brief description of the SDSSMOC4 data and an explanation of the sampling procedure are given in Sects. 2 and 3. The technique for resolving and defining the identified trends are given in Sections 3.3, while in Section 3.4 we add the Carvano groups to the new spatial configuration and discuss their placement in relation to the identified features and the scheme as a classification tool. The results are discussed in Section 4, along with the asteroid classification. We conclude and summarize our results in Section 5.

\section{Selection of sample from SDSS Moving Object Catalog}

The fourth release of SDSS Moving Objects Catalog is now the largest
photometric dataset. of small solar system bodies \citep{2002SPIE.4836...98I,2010PDSS..124.....I},
containing 471~569 detections of moving objects, where 202~101 are
linked to 104~449 unique objects from the ASTORB list \citep{1994IAUS..160..477B}.
The SDSSMOC is composed of a sample of asteroids and, in much smaller number, comets \citep{2010Icar..205..605S}. Moreover, the catalog may contain asteroid of sizes ranging from around 60 km (considering H=11.0 and geometric albedo of 5\%) to 700 meters (H=18.0 and geometric albedo of 25\%) and members of several families (\citet{2008Icar..198..138P}.

Observations obtained from 519 observing runs up to March 2007, 
using the  \emph{u', g', r', i'  z'} filter system \citep{1996AJ....111.1748F}
provide \emph{\emph{PSF magnitudes}} and the corresponding uncertainties at 355.7 nm, 482.5 nm, 626.1 nm, 767.2 nm, and 909.7 nm band centers and at bandwidths of 46.3 nm, 98.8 nm, 95.5 nm, 106.4 nm, and 124.8 nm, respectively.
The PSF magnitude of the sources is measured by fitting the point spread function model to the detected  objects
\citep{2002AJ....123..485S}.
Since the colors were obtained almost simultaneously, rotational
variations can be neglected for most of the observed asteroids. 
The SDSS magnitudes are in the asinh photometric system \citep{1999AJ....118.1406L},
which are convert to flux density\footnote{\url{http://ned.ipac.caltech.edu/help/sdss/dr6/photometry.html}} in order to derive the reflectances ($R_{j=\{u,r,i,z\}}$) corrected from solar contribution\footnote{\url{http://www.sdss.org/dr5/algorithms/sdssUBVRITransform.html}} and normalized at \emph{g'} band.

In what follows, all observations of undesignated asteroids have
been excluded, along with all detections 15 degrees from the Galactic
plane owing to including sources in crowded
stellar regions, and all observations with $|DEC|<1.26$ and negative galactic latitude, 
as recommended by the SDSSMOC authors\footnote{\url{http://www.astro.washington.edu/users/ivezic/sdssmoc/sdssmoc.html}}.
The resulting  SDSSMOC sample contains 103~627 detections
linked to 70~240 asteroids.In the following section, this sample was used to preliminarily characterize the spectral continuity and to select the best subsample for tracing their linear equation.

\subsection{Analysis of the sample}

\begin{center}
\begin{figure}[h]
\raggedleft
\includegraphics[scale=0.9]{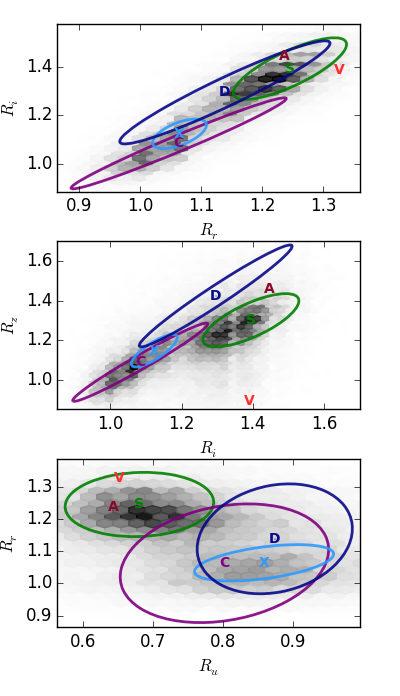}
\par

\caption{\label{fig:SDSSMOC_reflectances} SDSSMOC reflectances plotted
beside the taxonomic locus of SDSS asteroids with S3OS2 classification. In the background, there is a density plot with all 
SDSS observations}.
\end{figure}
\end{center}

 All the  major characteristics of asteroid spectra in the visible - presence of the 1 micron band, spectral slope, and drop in the UV - 
are sampled by the SDSS colors, and the relative importance of each characteristic is what ultimately defines the taxonomic classes (at least on the taxonomic schemes  exclusively based on spectral information).  These spectral characteristics induce correlations between the SDSS-derived reflectances.

 Figure \ref{fig:SDSSMOC_reflectances} shows
the plots of $R_{r}$ vs. $R_{i}$, $R_{i}$ vs. $R_{z}$, and $R_{u}$ vs. $R_{r}$ 
for the selected SDSSMOC sample. In the first plot, a general positive
correlation between the two variables is evident and reflects the
behavior of the spectral slope of the asteroid spectra. In the second plot, the
objects are more scattered, but some concentrations emerges overlapped
with the general positive correlation between $R_{i}$ and $R_{z}$, possibly
implying a finer structure of the behavior of the sample in the
reflectance space.  Two denser clusters of objects are also present
in the last plot, but $R_{u}$ and $R_{r}$ are not correlated.  

To better understand the observed behaviors, we superimposed some taxonomic
classes on these plots.  To do so, we selected a sample composed of those SDSS
asteroids that were previously observed spectroscopically and that have a
reliable taxonomic classification. We chose the S3OS2 survey, 
a spectroscopic sample obtained with uniform instrumental conditions \citep{2004Icar..172..179L},  
and we found 100 asteroids from this survey
that have at least one SDSS detection.  Using these "matching"
observations, we calculated (1) the variance-covariance matrix of their
four photometric variables and (2) the coordinates (and relative error
bars) in the space of the photometric variables of the center of the
taxonomic classes present in the sample.  In Figure
\ref{fig:SDSSMOC_reflectances}, each taxonomic type is represented by
an ellipse whose semi-axes  are defined by the eigenvectors of the variance-covariance matrix ($\pm 2\,\sigma$) 
plotted over the SDSSMOC sample. 

In the $R_{i}$ and $R_{z}$ plot, two different trends
emerge when the taxonomic classes are taken into account: the first
contains the taxonomic classes of asteroid with featureless spectra (B, C,
X, and D types), which have a slightly higher slope than that of the whole
sample; the other includes the taxonomic class  to which
asteroids presenting 1-micron absorption in their spectra  belong (A, S and
V types). However, the overlap of the taxonomic classes seen in Figure \ref{fig:SDSSMOC_reflectances}
indicates that the reflectances are not the best variable space for quantifying trends in the spectral features. 
Another problem here is also the contamination of the dataset by noisy
observations. In what follows, we transform the asteroid reflectance
to the more suitable set of variables and derive a new methodology
for obtaining the linear equation to observed trends.

\section{Methodology}\label{methodology}

\subsection{Principal component analysis}

  The issue of the best variable space can be addressed by applying principal component
analysis (PCA), which has been widely used in connection with asteroid taxonomy and which
constitutes the backbone of both Tholen and Bus taxonomic systems
\citep{1984PhDT.........3T,2002Icar..158..146B}. 
PCA have also played an important role in the analysis of SDSS photometry before \citep{2002SPIE.4836...98I,2005Icar..173..132N,2006Icar..183..411R}.

The PCA rearranges all data placement and dimensionality by
characterizing it through their principal correlated axes  (e.g., \cite{Press:1988:NRC:42249} and \cite{PCAJolliffe}). 
The principal components (abbreviated to PC for representing the variables and not the method) carry different amounts of sample variance, usually connected to
information contained in data, allowing correlations in the data to become more pronounced.
The PC were computed
by singular value decomposition \citep{2002physics...8101W} method using the Scikit-learn package for the Python programming language \citep{scikit-learn}. We applied it with whitening and
kept the same number of variables. The whitening is a normalization that
rescales the data by dividing the component matrix by each PC standard
deviations, thereby maximizing the output for real correlations and equalizing
the variables. Whitening does not alter real correlations found in the
data, but only improves visualization. Then, the equation of
transformation of reflectances to PC is given by

\begin{equation}
\vec{PC_{j}}=\frac{E*\vec{X_{j}}}{\vec{S}*\sqrt{N}}
\end{equation}where $\vec{X_{j}}$ is a SDSS observation, in reflectance; $E$ is the
eigenvector matrix, containing the weight of each reflectance
on the definition of each PC in each row; $\vec{S}$ is the standard
deviation vector; and $N$ is the total sample size. The transformation
gives us a new four-dimensional variable $\vec{PC_{j}}$.

We can now use the distribution of the principal components to visualize the relation between spectral characteristics
and the SDSS reflectances
better and to analyze how this is affected by
uncertainties in the data.

\subsection{Probing the role of the uncertainties in the SDSSMOC}

\begin{center}
\begin{figure}[h]

\begin{centering}
(a)\includegraphics[scale=0.5]{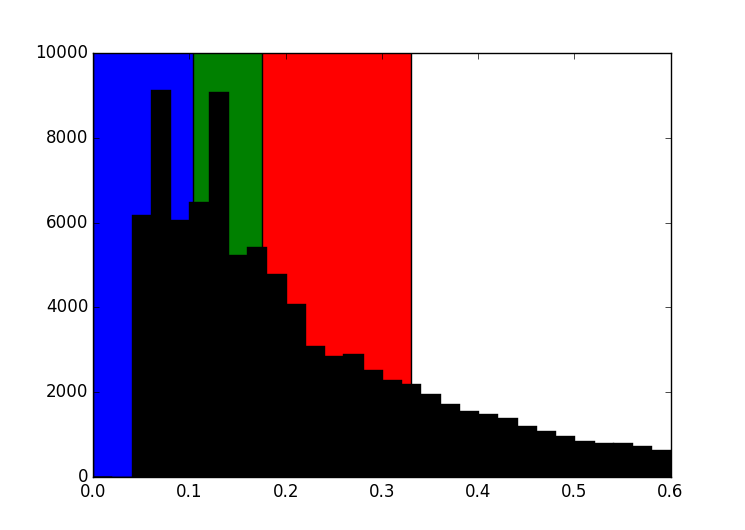}
\par\end{centering}

\begin{centering}
(b)\includegraphics[scale=0.45]{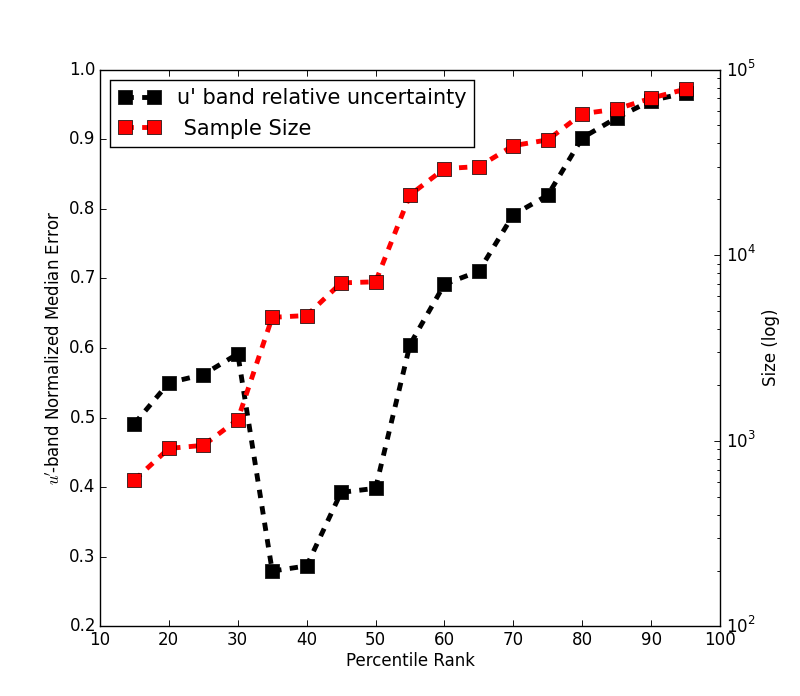}
\par\end{centering}

\caption{\label{fig:Ranks} (a) TDistribution of errors in the u'-band. 
The blue, green, and red backgrounds denote the Q1, Q2, and Q3, respectively.
(b) Plot of the \emph{u'}-band normalized median error and sample size according to percentile ranks.}
\end{figure}

\par\end{center}

\begin{center}
\begin{figure*}[h]

\begin{centering}
(a)\includegraphics[scale=0.6]{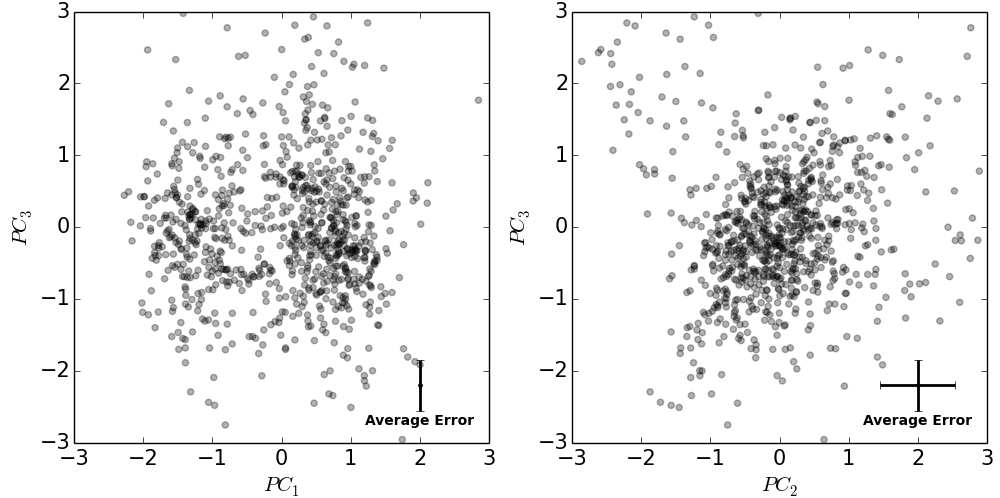}
\par\end{centering}

\begin{centering}
(b)\includegraphics[scale=0.6]{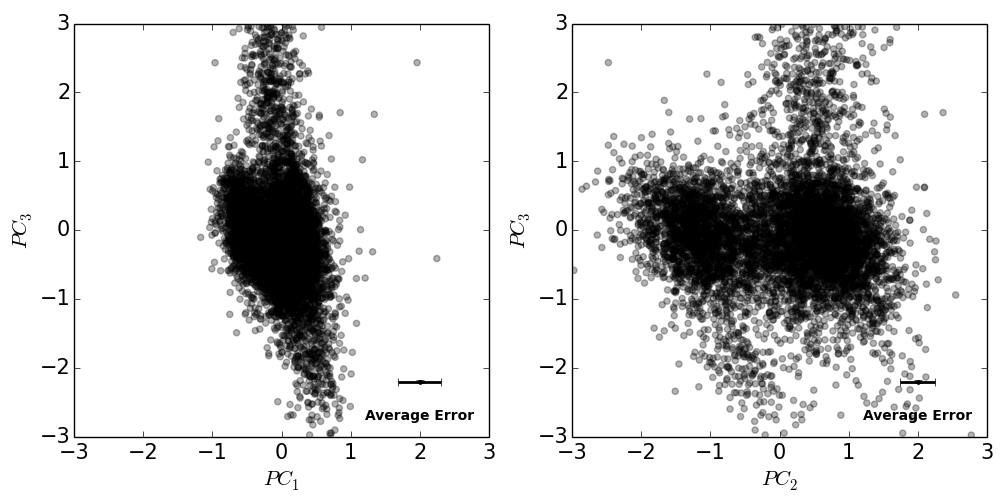}
\par\end{centering}

\begin{centering}
(c)\includegraphics[scale=0.6]{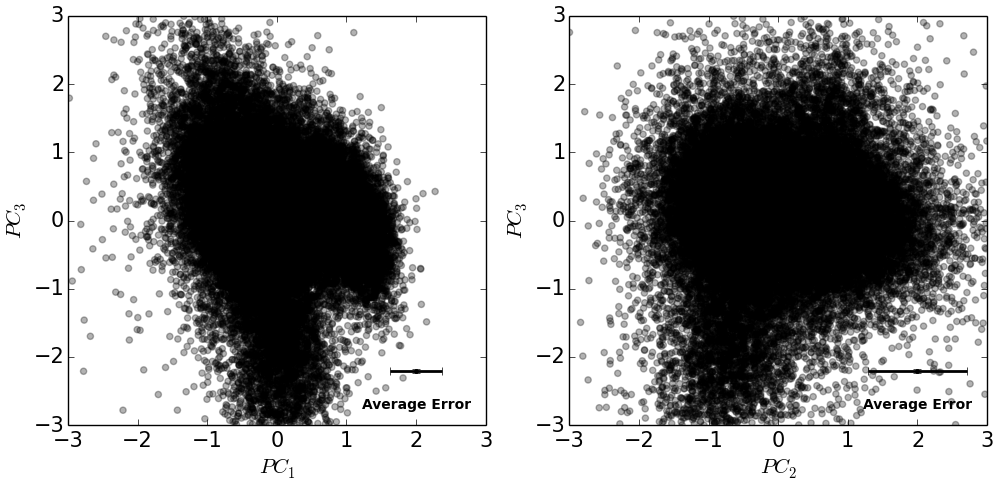}
\par\end{centering}

\caption{\label{fig:Quartiles_PCA}PC2 vs. PC1 and PC3 vs PC2 for (a) 1st (Q1), (b) 2nd (Q2), and (c) 3rd quartile (Q3) samples. 
 The principal components of each sample have been computed independently. The error bars, contained in the bottom right corner, represent the average uncertainty in the two axes.}
\end{figure*}

\par\end{center}

  The SDSSMOC contains observations of moving objects taken at different observational configurations, atmospheric conditions, stellar background, and position on the CCD. All these variables contribute to the uncertainties, or noise, of the data, which might reflect on the observed spectral continuity. Thus, to understand the effect of noise on the data set, 
the SDSSMOC sample was selected at three standard percentile ranks disstinguished in the distribution of uncertainties of the four filters. The percentile rank defines how much, in percentage points, of the data is found under an upper error limit.
This kind of progressive sorting allows visualizing and analyzing the increase in the noise weight on the SDSSMOC sample.

 We then used quartiles of the uncertainty distribution containing 25\%, 50\%, and 75\% of the sample to define three subsamples: Q1, Q2, and Q3.
(a given example in Figure \ref{fig:Ranks}a presents the three sampling in the \emph{u'}-band).
Table \ref{tab:sample-data}  gives, for each SDSS filter, the upper error limits and sample size of Q1, Q2, and Q3, as well as their total standard deviation, total median uncertainties, and completeness. The completeness is the magnitude limit where the reliability of point sources to be detected drops to 95\%, and the photometric measurements has lower quality\footnote{\url{http://www.sdss3.org/dr8/imaging/other_info.php}\#completeness}.

In Figure \ref{fig:Quartiles_PCA}, we plot PC3 vs PC1 and PC3 vs PC2 for the three quartiles, and it is clearly shown how the increase in the noise blurs the information contained in the data and the increase in size modifies the PC calculated independently for each sample.
In PC3 vs PC1 of Q1, the data distribution resembles what is observed by \citet{2007A&A...467..749W}, which used 1332 SMASSII spectra to derive a taxonomic classification for \emph{Gaia} photometric data. In our plot, two groups with elongated shapes appears. These two groups are related to spectra with and without absorption band (C and S). The remaining points are too sparse to derive any conclusion about their distribution.  The PC3 vs PC2, on the other hand, is dominated by the errors that blur the separation between the clusters of featureless spectra and spectra with the one-micron band.

 In the PC space of Q2, it is possible to visualize two clear concentrations, spreading in the PC3 direction. These concentrations, which are better described as trends, are best resolved in the PC3 vs PC2 space.  It is noteworthy that although PC3 contains the largest spread, the largest part of the error is contained in PC1. Therefore, that these two concentrations are not unimodal outside the C and S groups indicates the existence of a continuous distribution of spectral types.
 
 Finally, in Q3, the sample is highly dominated by uncertainties, and the C and S groups are barely distinguished from each other, although the trends are still visible. Moreover, there are no large differences between the eigenvector matrix of the Q2 and Q3, which indicates the robustness of these correlations in the data ,which are unchanged even when the data quality decreases.

 To select the sample that discerns the trends better, we then sampled the SDSSMOC uncertainties in several percentile ranks and computed the normalized median error for each SDSS filter. This quantity measures the relative importance of the formal photometric errors in the dispersion of the data. The result, when sampling from ranks of 15\% to 100\%, is that the normalized median error in $R_{u}$, $R_{r}$, $R_{z}$, and $R_{i}$ vary from 25-82\%, 7-18\%, 10-15\%, and 9-27\%, respectively. The \emph{u'}-band normalized median error has the largest variation, because this band has the lowest quantum efficiency among the four SDSS filters at five times less than the \emph{r'}  band \citep{1998AJ....116.3040G}.
Nonetheless, we prefer not removing the \emph{u'}-band, because it is the sole filter probing the near ultraviolet, and it brings important information, since in this spectral region, owing to charge transfer inside iron-bearing minerals, an absorption band may be present, causing a significant drop in the spectra.  

 Therefore, we decided to analyze the quality of each sample relative to the \emph{u'}-band normalized median error. This behavior is shown in Figure \ref{fig:Ranks}b, where three regimes of data sorting can be perceived. If the sampling is too restrictive, with scores below 30\%, the sample size is too small and dominated by errors, as is the case for Q1 (Figure \ref{fig:Quartiles_PCA}a). If the sampling is loose, increase in sample size correlates to the increase in noise, which means that the sample is also dominated by errors. Therefore, only between scores of 30\% to 50\% do we find a minima in the \emph{u'} relative uncertainty and a near plateau in sample size.
On the basis of these considerations, we decided to continue our analysis using Q2, since it offers the best ratio between uncertainty and size and keeps the trends clearly separated.

\begin{center}
\begin{table*}[t]
\caption{\label{tab:sample-data}Characteristics of the SDSSMOC samples. {All samples are normalized to the g' filter, 
where the completeness is 22.4 mag.} }

\centering{}
\begin{tabular}{cccccccc}
\hline
{\bf \emph{filter}} & {\bf standard deviation} & {\bf total average error} & {\bf Q1} & {\bf Q2} & {\bf Q3} & {\bf Completeness (95\%)}& \tabularnewline
\hline
\hline
\emph{u}  & 0.213 & 0.126 & 0.103 & 0.180 & 0.330 & 22.1 & \tabularnewline
\emph{r}  & 0.178 & 0.025 & 0.026 & 0.026 & 0.033 & 22.1 & \tabularnewline
\emph{i}  & 0.219 & 0.027 & 0.026 & 0.033 & 0.041  & 21.2 & \tabularnewline
\emph{z}  & 0.226 & 0.048 & 0.041 & 0.076 & 0.134 & 20.3 & \tabularnewline
\hline
 & & Total sample size (number of observations) & 905 & 7,190 & 41,820 &  & \tabularnewline
\hline
\end{tabular}
\end{table*}

\par\end{center}

The eigenvector matrix ($E$) and $\vec{W}$
weight vector for SDSSMOC-Q2 are given by
\begin{equation}
E=\left[\begin{array}{cccc}
0.228 & 0.296 & 0.289 & 0.286\\
-1.209 & 0.206 & 0.505 & 0.2378\\
0.106 & 1.232 & 0.750 & -2.119\\
1.075 & -4.762 & 4.973 & -0.955
\end{array}\right]
\end{equation}

\begin{equation}
\vec{W}=S^{2}/N=\left[\begin{array}{c}
0.82\\
0.14\\
0.04\\
0.005
\end{array}\right]
.\end{equation}

 Given vector $\vec{W,}$ then $\vec{S}$ can be easily derived. Vector $\vec{W}$
represents the variance weight ratio of each PC. The PC values of each SDSS observation can be computed by inserting E and S into Equation (1). For Q2, N=7190.
The slope of the asteroid spectra is reflected in the large amount of variance contained in PC1, 
while PC4 can be neglected since it holds less than 1\% of the total variance. 
The space PC2-PC3 will thus be used since it is the most representative.

The variables for the chosen subset are strongly linearily correlated as
indicated by the 82\% of variance accounted by PC1, which also contains
an equal proportion of each SDSS band, implying that it is
intrinsically connected to the spectral slope \citep{2005Icar..173..132N}. PC2 weighs on
$R_{u}$ and $R_{i}$, and the negative sign of the score of
$R_{u}$ means that when spectral slope is high, the
ultraviolet drop is generally deeper. PC3 correlates with $R_{z}$, the variable
containing the information on the one-micron silicate absorption
band. Therefore, trends in the spectral features that are relevant
to classification can now be discerned in PC space. To use this
characteristic, we need to obtain equations of the axis that define
the trends in the PC space.

\subsection{The linear equations for the spectral trends in PC space}

 The next step in our analysis will be to obtain the linear equations for the trends that can be used for classification and quantification purposes.  To perform this, we divided the observations into regions and replaced them by vector, where the only information kept was their position and their direction. Then, we clustered them according to their vicinity and common alignment. This was performed by applying a multidimensional histogram tool from Numpy
\footnote{\url{http://docs.scipy.org/doc/numpy/reference/generated/numpy.histogramdd.html}} and Scipy.hierarchy tools \footnote{\url{http://docs.scipy.org/doc/scipy/reference/cluster.hierarchy.html}}. 
The procedure is summarized in the following steps:

\begin{enumerate}
\item The SDSSMOC-Q2 is partitioned into four-dimensional cells, were each PC
is divided into seven regions ($7^{4}$ = 2401 cells). 

\item Cells containing over 15 observations had their PC median ($\vec{x}_{i}$)
and variance-covariance matrix recorded. The median gives the center
of the cells, while the first eigenvector ($\vec{e}_{i}$) of variance-covariance
matrix indicates the main direction of correlation of each cell.

\item The distance among all cells are calculated through the equation:
\[
d_{ij}^{2}=||x_{i}-x_{j}||^{^{2}}+||\hat{e_{i}}-\hat{e_{j}}||^{2}.
\]
The first eigenvector is included in the distance equation to account
for cells of similar orientation.

\item The cells are hierarchically arranged according
to their distances as computed above. Similar cells are organized
by pairs until reaching a binary tree structure \citep{1979IEEE...4B}.
The dendrogram in Fig. \ref{fig:BinaryTree}a shows how the sectors
are coordinated. The nodes are points where pairs met, representing
the pairs below and forming a supercluster. Inspecting the nodes
allowed us to search for correlation features among the observations,
such as the observed trends.

\item Two nodes (I and IV, Figure \ref{fig:BinaryTree}b) contain most of the samples belonging to
the trends recognized in the previous analysis. These nodes form
superclusters that have the median and variance-covariance matrix
calculated from all observations composing it. From the variance-covariance
matrices, we computed the four eigenvectors that gives four main axes
of each supercluster.
The first eigenvector is the main axis, and it represents the orientation and
direction of the trends. Then, from the main axis we obtain the linear
equation for each trend in the full PC space:
\end{enumerate}

\begin{align*}
\emph{Red}:\,\,\,0.0091\cdot PC1+0.0190\cdot PC2-0.0056\cdot PC3 &-0.0226\cdot PC4\\
& +0.0017=0
\end{align*}

\begin{align*}
\emph{Blue}:\,\,\,0.0683\cdot PC1-0.108\cdot PC2-0.0636\cdot PC3 &+0.1032\cdot PC4\\
& -0.2267=0.
\end{align*}

From now on, we decided to address the one-micron band featured and featureless trends to Red and Blue, because of the colors chosen to represent them in Figure \ref{fig:BinaryTree}b.

These equations allow us  to compute the observation's distance ($d$) from and position ($z$) for each trend. The projection of the lines describing the two trends on the PC3-PC2 plane are given in Figure \ref{fig:BinaryTree}b. 
The position in the main axis of one trend quantifies the intensity of the related spectral feature (band depth or spectral slope/UV-drop) and is calculated by the scalar product of the main axis vector and the observation vector $PC_{j}$.
An observation is regarded as belonging to one trend if its distance with respect to each trend center 
satisfies the following condition:

\begin{equation}
\vec{d}_{red}=\left[\begin{array}{cccc}
-0.1195 & -0.5328 & -0.0056 & -0.8377\\
-0.0944 & -0.8325 & -0.0518 & 0.5434\\
0.9854 & -0.1472 & 0.0716 & -0.0474
\end{array}\right]\left[\begin{array}{c}
PC1-0.204\\
PC2-0.590\\
PC3+0.154\\
PC4-0.694
\end{array}\right]_{obs}
\end{equation}

\begin{equation}
\vec{d}_{blue}=\left[\begin{array}{cccc}
-0.2306 & 0.2464 & -0.0755 & -0.9382\\
0.0324 & 0.8577 & 0.4810 & 0.1786\\
0.9263 & -0.1086 & 0.2335 & -0.2750
\end{array}\right]\left[\begin{array}{c}
PC1+0.147\\
PC2+1.201\\
PC3+0.407\\
PC4-0.788
\end{array}\right]_{obs}
\end{equation}

\[
||\vec{d}_{red\, or\, blue}||<1,
\]

where the three rows of the matrix are given by the secondary eigenvectors characterizing each trend.
It is important to note that one observation may belong to a single trend or to both. 

\begin{center}
\begin{figure*}[p]
\begin{centering}
\begin{tabular}{>{\centering}b{12cm}c}
 & \tabularnewline

(a)\includegraphics[scale=0.65]{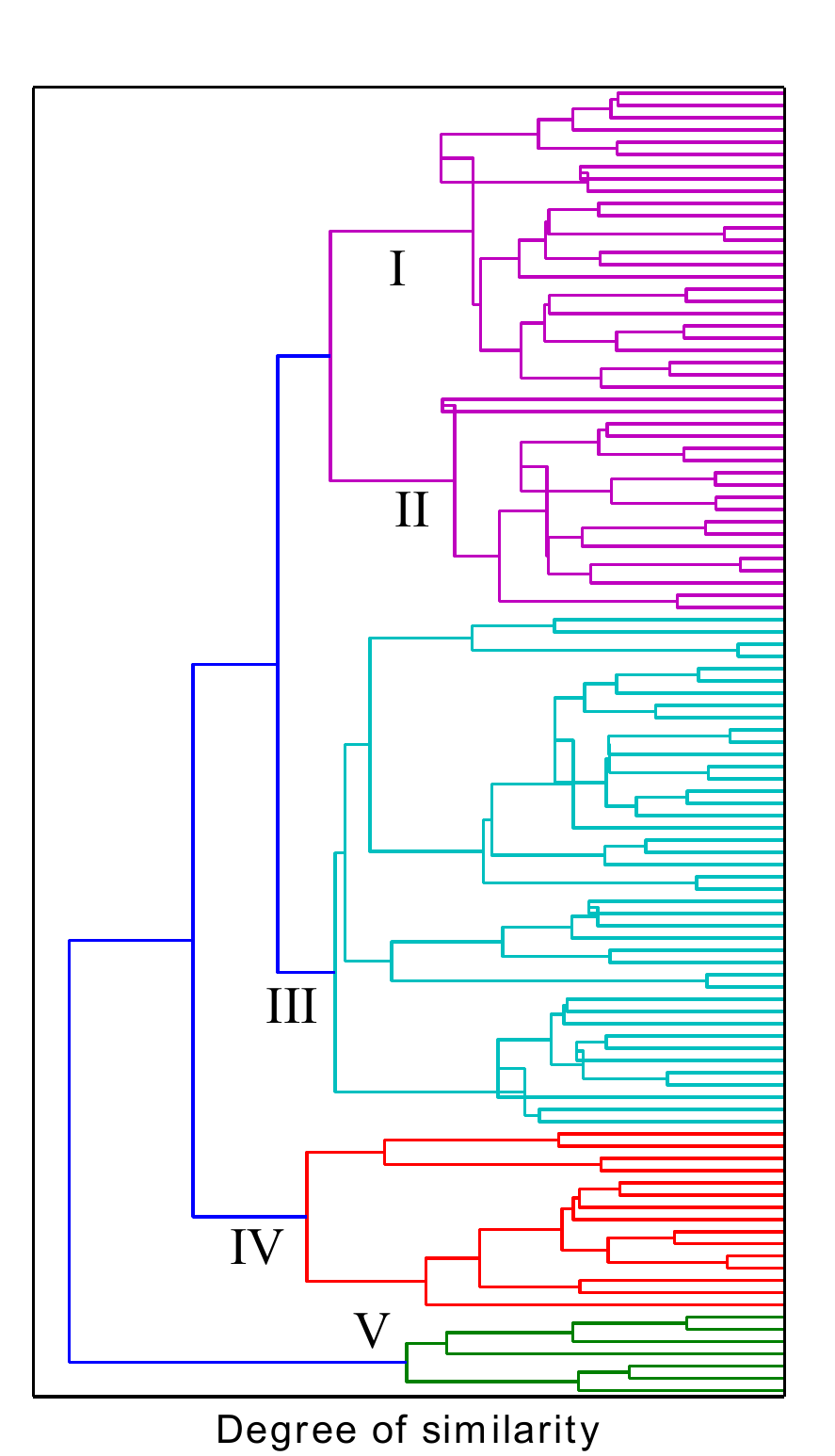} & \tabularnewline
(b)\includegraphics[scale=0.6]{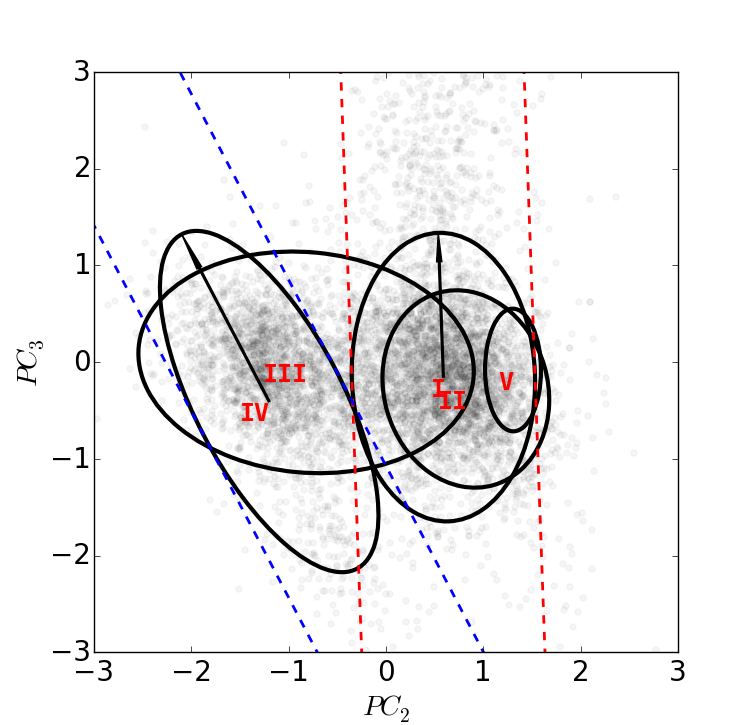} & \tabularnewline
\end{tabular}
\par\end{centering}

\caption{\label{fig:BinaryTree}
(a) Binary tree dendrogram. Each superior node supports several lower pairs. Different colors highlight the biggest nodes.
(b) Trends. The ellipses are the representation in PC space of the nodes of same roman letter in Figure (a). The nodes I and IV represent the equation for the Blue and Red trends. 
The contours are the projected ellipsoids. The dotted lines are the projected cylinder obtained through the secondary eigenvectors of the variance-covariance matrix of the I and IV nodes.}
\end{figure*}

\par\end{center}

\subsection{Locus of asteroid classes}

 Once the trends are defined in the PC2-PC3 plane,
we can relate them to the asteroid taxonomic classification and look where the groups are
located in this PC space. The superposition of the taxonomic classes
to the PC2-PC3 plot will help in interpreting the meaning of the two
trends in terms of the spectral behavior of the asteroid.  We decided
to use the photometric classification available in the literature
\citep{2011PDSS..145.....H} to which we add the B group (not included in the original work) 
using SDSSMOC observations spectroscopically
classified as B and F in \citet{2010PDSS..123.....N}.  We applied the
same procedure as described in section \ref{methodology} to insert these
taxonomic groups in the PC space. The Figure
\ref{fig:Carvano} shows the 2$\sigma$ locus of each class overlapped
with the distribution of Q2. 

 The center of the classes in the PC space can now be projected
into both axes defining the trends. For each class in each trend, Table \ref{tab:Trends}
gives its distance and position, as defined in Section 3.3.
The values of the z-distances along each trend
were shifted so that the S class is at the origin of the \emph{\emph{Red}} trend
and the C class is at the origin of the \emph{\emph{Blue}} trend.
The position of the class in each trend and their possible
interpretations are discussed further in the following section. 

\begin{center}
\begin{figure}[h]
\begin{centering}

\par\end{centering}

\begin{centering}
\includegraphics[scale=0.4]{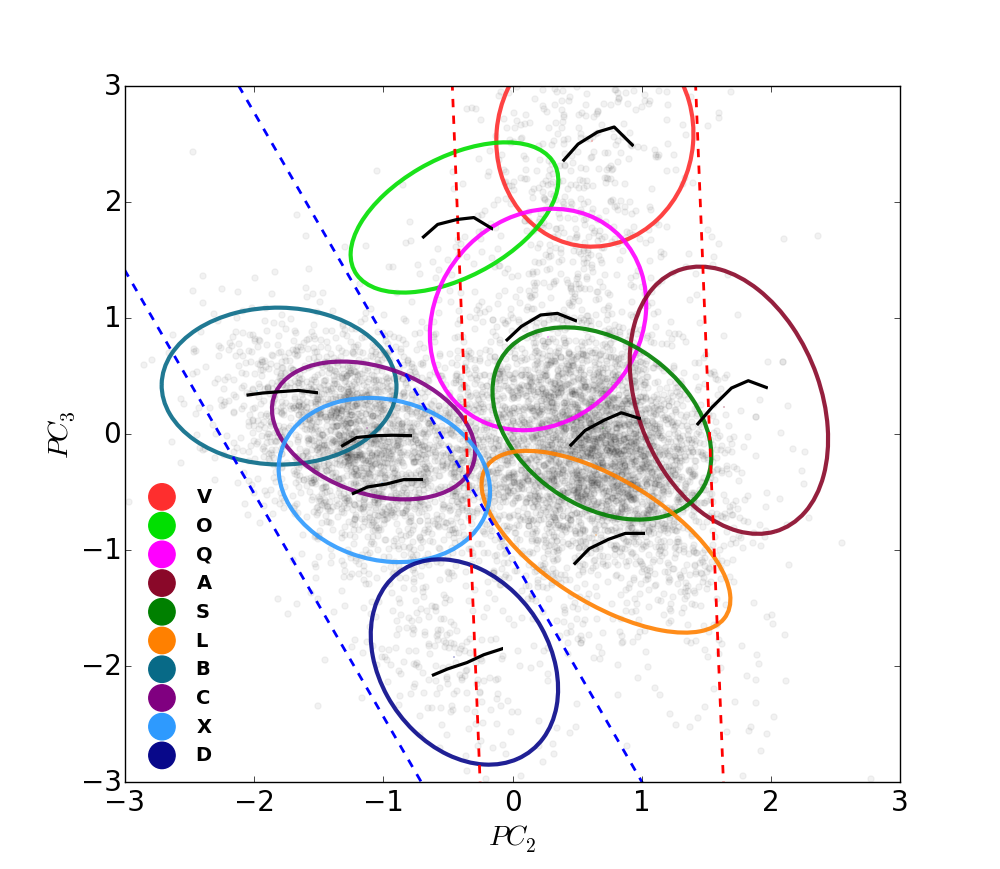}
\par\end{centering}

\caption{\label{fig:Carvano} Loci of the taxonomic classes in PC3 vs
PC2 space, represented by ellipses where the axes represent $2\sigma$ 
of the distributions. The $2\sigma$ confidence region for the Blue and the Red trends is
indicated by the dotted lines.  }
\end{figure}

\par\end{center}

\begin{center}
\begin{table}[h]
\caption{\label{tab:Trends} Distance and position of the center of the taxonomic groups on the trends. The intervals are given in 1$\sigma$.}

\begin{centering}
\begin{tabular}[t]{crrrr}
\hline
class &        $d_{blue}$              &          $d_{red}$              &           $z_{blue}$                &      $z_{red}$                    \\
\hline \\
 V    & 2.792 $^{2.96}_{2.44}$   & 0.146$^{0.44}_{0.12}$     & 1.226$^{1.66}_{0.82}$      & 2.496$^{2.96}_{2.02}$     \\
 \\
 O    & 1.577$^{2.01}_{1.47}$    & 1.207$^{1.26}_{0.88}$     & 1.382$^{1.43}_{1.14}$      & 2.066$^{2.30}_{1.89}$    \\
 \\
 Q    & 1.684$^{1.75}_{1.31}$    & 0.413$^{0.69}_{0.18}$     & -0.106$^{0.08}_{-0.55}$    & 0.808$^{0.36}_{0.88}$    \\
 \\
 A    & 2.679$^{2.99}_{2.55}$    & 1.097$^{1.40}_{1.09}$     & -1.431$^{-1.14}_{-2.01}$   & 0.107$^{0.19}_{-0.47}$     \\
 \\
 S    & 1.673$^{1.98}_{1.56}$    & 0.000$^{0.35}_{0.15}$     & -1.009$^{-0.89}_{-1.47}$   & 0.000$^{0.24}_{-0.35}$    \\
 \\
 L    & 1.165$^{1.25}_{0.91}$     & 0.102$^{0.57}_{0.13}$     &-1.772$^{-1.23}_{-1.88}$    & -0.918$^{-0.85}_{-1.10}$   \\
 \\
\hline \\
B     & 0.707$^{1.12}_{0.73}$     & 2.717$^{3.15}_{2.70}$     & 0.601$^{0.79}_{0.31}$      & 0.451$^{0.58}_{0.11}$   \\
\\
C     & 0.000$^{0.39}_{0.17}$     & 1.951$^{2.30}_{2.72}$     & 0.000$^{0.31}_{-0.41}$    & 0.115$^{0.29}_{-0.26}$  \\
\\
X     & 0.229$^{0.41}_{0.20}$     & 1.763$^{1.75}_{1.39}$     & -0.554$^{-0.24}_{-0.87}$  & -0.383$^{-0.02}_{-0.55}$  \\
\\
D     & 0.401$^{0.72}_{0.10}$     & 1.220$^{1.52}_{0.82}$     & -2.233$^{-2.22}_{-2.73}$  & -1.938$^{-1.88}_{-2.01}$  \\
\\
\end{tabular}
\end{centering}
\end{table}
\end{center}

\section{Discussion}

 In the scheme represented by the Figure \ref{fig:Carvano}, the Blue contains the asteroid belonging to the
C$\rightarrow$X$\rightarrow$D classes characterized by featureless
spectra, while the \emph{\emph{Red}} trend refers to asteroid classified as
L$\rightarrow$S$\rightarrow$Q$\rightarrow$V types that have absorption
features in their spectra.  As expected, C and S classes gather most
of observations for each trend, while all the other classes seem to "leak" out
from them.  Following the B$\rightarrow$D direction in the Blue trend, 
the spectral slope increases as a consequence of the PC2 along
the trend, corresponding to a deeper and deeper drop in the \emph{u'}-band
reflectance.  The "red" trend vertically crosses the PC2-PC3
plane at quasi constant values of PC2 in the direction
L$\rightarrow$V. There is clearly a spectral continuity from spectra
with a higher slope bearing hints of a 1-micron shallow
band to spectra with lower slopes and deeper bands.  The O and A loci
fall partially out of the confidence region of the Red trend and,
together with the Q-types locus, form a O$\rightarrow$Q$\rightarrow$A
sequence, almost parallel to the "blue" trend.
 This sequence has its change in spectral slope and \emph{u'}-band drop.
They have similar evolution to the \emph{\emph{Blue}} trend, 
where both features increase from O$\rightarrow$A, so much like the
C$\rightarrow$D sense.

 It is noteworthy that there are diverse loci that are largely superimposed, in particular, 
 C and X, S and L, and S and Q. 
Such superimposition is observed in all the PC space and is not an effect of projection.
These classes have differences that are based solely on the varied strength of their spectral features, mostly spectral slope and depth of silicate band,
which we have demonstrated are continuous. 

This is therefore a direct consequence of the adopted reference taxonomy and the lack of
some of these criteria when a different set of variables is used.  In
our representation what is important is that
spectrophotometric data have an internal coherence, that is to say that a
continuum of the spectral behavior of the asteroids of the SDSSMOC is maintained. 
In fact the aim of our analysis is to discuss
and interpret the structure of that continuum and to use it to better
characterize the analyzed population, as well as to gauge observational effects (i.e., phase reddening), which may modify the specta of asteroids.
This kind of approach has been done on much smaller data
sets by \citet{1986LPI....17..985B} and by \citet{1987Icar...72..304B}.

Bell collapsed the Tholen taxonomic classes into three
superclasses (primitive, igneous, metamorphic), thereby accounting for the degree
of metamorphic heating the asteroid has undergone. Barucci et al.
propose four evolutionary trends to interpret the asteroid composition nature through a
taxonomy scheme based on the ECAS \citep{1985Icar...61..355Z}
and the IRAS albedo catalog. These trends account for i) the volatile
content reduction, ii) the evolution of the primitive solar nebula
condensate toward enstatite silicates, iii) the increasing of the
differentiation, and iv) the effect of the collision and fragmentation of
differentiated bodies. Since that work, however, 
it has been realized that there are several factors that may influence the shape of an asteroid spectrum that is not directly related to composition. 
Two examples are space weathering \citep{1996Icar..122..366M,2010Icar..209..564G,2013A&A...554A.138L,2014AAS...22411911M} and phase reddening \citep{2012Icar..220...36S}.


The Blue trend accounts for the dark primitive
asteroids. Increasing values of PC3 could be interpreted in terms of variation in the bulk
chemical and mineralogical composition from the more primitive (D
type) toward the more evolved dark B-type objects 
\citep{2010AJ....140..692Y, 2014A&A...568L...6P}, which could also
account for a decrease in volatiles or organics present on their surfaces. However, similar effects could also arise from space weathering or phase reddening. Also, spectra of asteroids with compositions similar to enstatite chondrites and achondrites and metallic meteorites would also fall in this blue trend (i.e., \cite{2003Icar..161..356C}). Finally, asteroids with compositions dominated by olivine-pyroxene whose bands have been suppressed for any reason (i.e., \cite{2011Icar..216..184R} and \cite{2014Icar..237..116R}) would also plot on the Blue trend.
  
The Red trend describes the behavior of the spectra of asteroids dominated by a olivine-pyroxene composition. Increasing values of PC3 could indicate their difference in the silicate compositions that are characterized by a progressive increase in the amount of pyroxene-olivine going from L to V types. Nonetheless, phase reddening and space weathering
also affect the PC3 by changing the spectral slope and band depth, which moves the asteroid spectrum in the trend's direction. 
Inside the confidence band of both trends, not much can be said about variability on the PC2,
where the errors have a major influence.
Objects having more extreme spectral characteristics (such as those of A or O asteroids) occupy the
border of the trend's band of confidence. 

\subsection{Asteroid classification}

 One application of the methodology developed in the previous section is to classify the spectra of asteroids according to their position inside the trends. 
To this effect we gathered the complete SDSSMOC sample of 202~101 detections and 
classified using the procedure described in the section 3.3. 
We obtain a classification of 139~233 observations corresponding to 79~647 asteroids, 
where 39.85\% belongs to Blue/featureless trend and  54.05\% to the Red/featured trend. 
Only 5.85\% of the observations are contained in the region intersected by both trends.
Although carrying less information  about the  shape of spectra when compared to other proposed classification schemes, this classification can be applied to a considerably larger number of observations.  In comparison with the previous SDSSMOC classification by \cite{2010A&A...510A..43C}, we find a n
increase of 29.5\% in the capability of classifying asteroid observations, which could not be classified by the authors. This represents 31~767 new observations being classified. 



\begin{center}
\begin{figure}[H]
\begin{centering}

\par\end{centering}

\begin{centering}
(a)\includegraphics[scale=0.4]{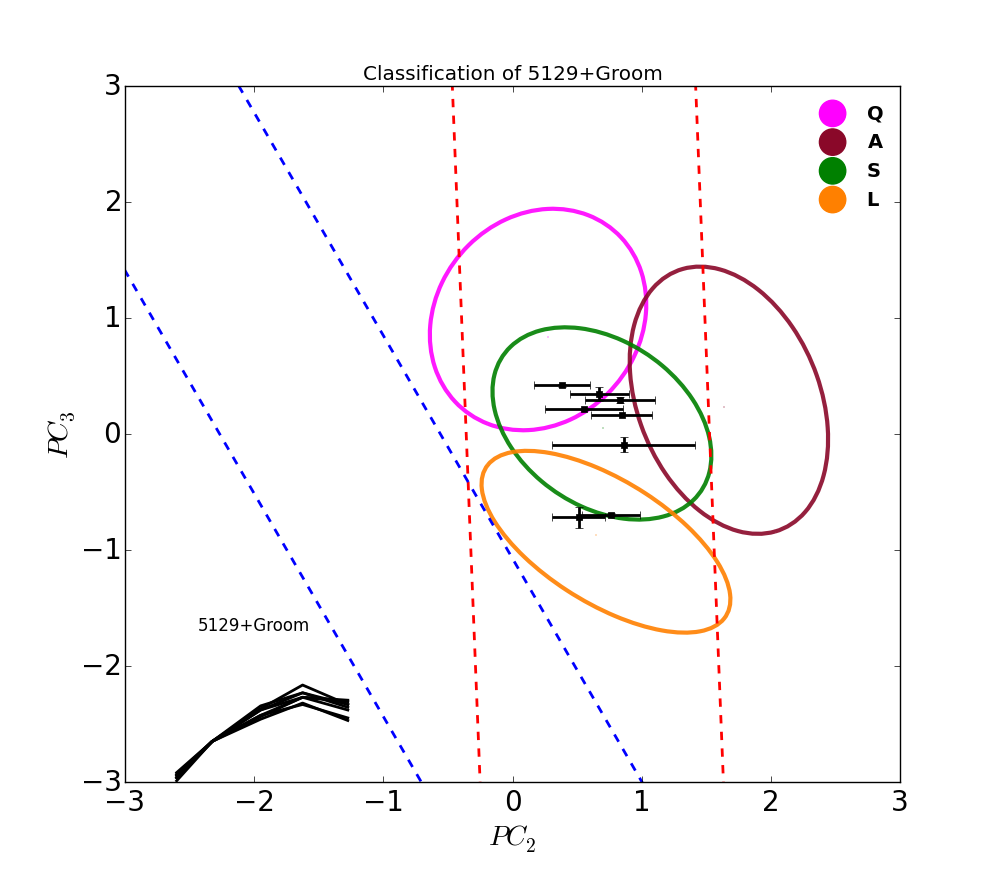}
\par\end{centering}

\begin{centering}
(b)\includegraphics[scale=0.4]{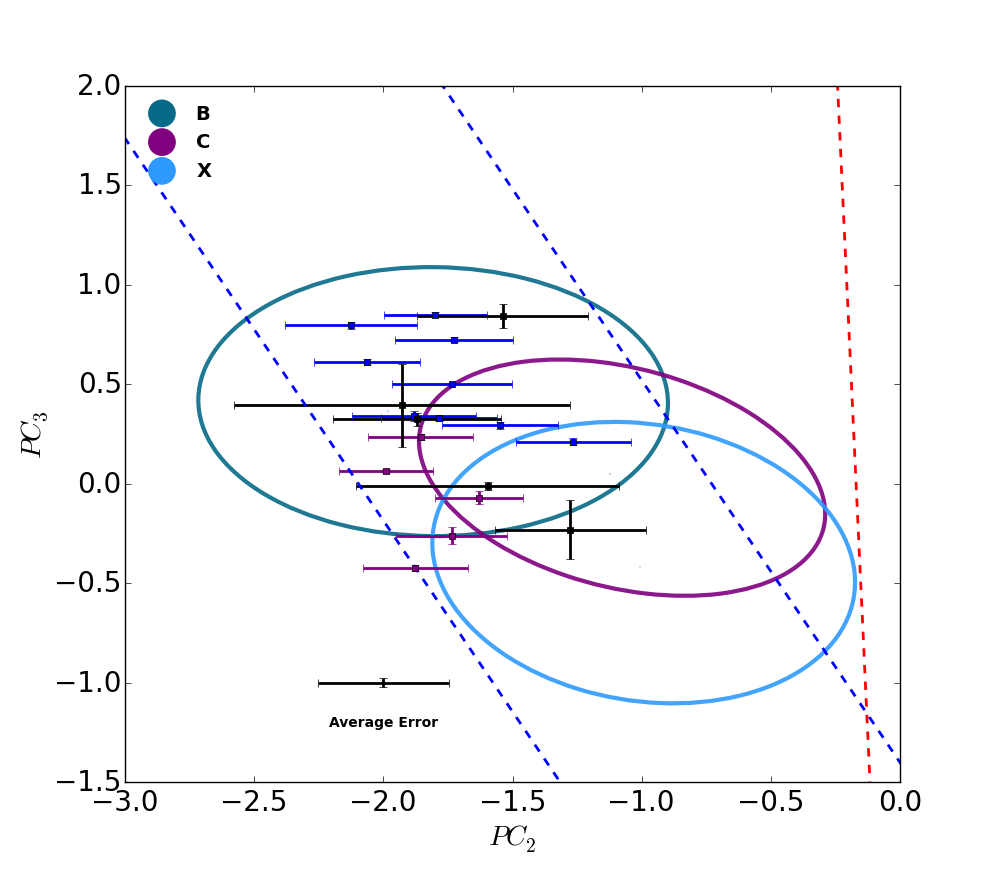}
\par\end{centering}

\caption{\label{fig:Application}  (a) Eight SDSS observations of asteroid (5129) Groom plotted in the scheme proposed. 
The error bars represent the errors transformed to PC space. The spectra of the observations are plotted together in the lower corner of the graph. The specific SDSS observation IDs are s14307, s0e2e7, sca267, sc8e1d, scf0dc, scdde4, scc719, and scbc65. (b) Three asteroids with multiple observations presenting spectral variability. The SDSSMOC observation IDs for each asteroid: (2252) CERGA (\textcolor{blue}{blue}): s3bafc, s3f6f3, s3eea4, s41189, s424bf, sbbb68, sf0125, sef985, sef5d2; (3631) Sigyn (\textcolor{purple}{purple}): s16290, s16001, sf0e38, s042fc, s132c2; and (55926) 1998 FE60 ({\bf black}): s17150, s18825, s17c18, s3924d, s2cd9b.}
\end{figure}

\par\end{center}

 A second application of the developed methodology is to provide an alternative way to classify asteroids on SDSS-based taxonomies.  This can be accomplished by calculating the PC using equations (1)-(3) and then plotting the coordinates in a diagram like Figure  \ref{fig:Carvano}, along with the boundaries of each class. Besides this visual classification is also possible to calculate the distance to the center of the taxonomic groups. These distances are also a quantitative indicator of how much the spectra of the observation deviates from each class template, which corresponds to the center each group. This allows for more efficient automated searches for specific spectral characteristics.

 As example of one application of our scheme, we selected one asteroid, (5129) Groom, observed eight times by the SDSS. 
In \cite{2011PDSS..145.....H}, (5129) Groom is classified as $S_{p}$, but one observation is Lp, and the other one is $S_{p}$/$L_{p}$.  
When we plot these observations in Figure \ref{fig:Application}a, we find them contained inside the Red trend, 
although spanning three groups:  L, S, and Q. 
Using the scheme proposed here, we are thus able to verify that the observations increase in their $z_{red}$  from -0.571, so well inside the L group, up to 0.591, at the bottom of the Q group. his variability cannot be attributed to phase reddening, owing to the small phase angle range (from 4.5 up to 11 degrees) of the diverse observations, and might be related to compositional differences  across the object surface.

 We may extend this application and verify asteroid spectral variabilities inside the boundaries of the taxonomic groups. 
We present in the Figure \ref{fig:Application}b, for example, three asteroids with multiple observations in the $B$ group: 
(2252) CERGA (9 obs.), (3631) Sigyn (5 obs.), and (55926) 1998 FE60 (5 obs.).
All these asteroids present spectral variablity that exceeds their intrinsical errors and the average uncertainty of all SDSSMOC-Q2 sample.
The variability is substantial not only along the PC3, but also along the PC2, as in the case of the asteroid (2252) CERGA.

\section{Conclusions and results} 

Using a subset of the visible reflectances from the
SDSSMOC catalog, we found that asteroids follow mainly two continuous
trends that describe their spectral behavior.  These are barely visible
in the reflectance space, but emerge clearly when data are translated
into the PC space. Information drawn from the
adopted dataset is contained within a band of the
PC2-PC3 plane defined by the direction of each trend.
These trends represent the evolutionary track of the two main characteristics of asteroid spectra:
the presence and depth of the 1-$\mu m$ band and the spectral slope.

 It is noteworthy that most of the classified SDSSMOC data belongs to one of the two trends. 
A total of 29.5\% more of the objects have been classified in \cite{2010A&A...510A..43C}.
This demonstrates that some objects in between the formal taxonomic classes are taken into account.
The given equations for the lines representing the trends, allows any object to be classified (asteroid, meteorite, or comet) for which the \emph{u'g'r'i'z'} photometry is available.

As a data product resulting from this work, we applied the PC
transformation given in Equation (1) to all observations of the SDSSMOC, and then used the equations 4 and 5 to calculate the distances of each observations along the axis that describe each trend. 
These lists will be made available as a product in the NASA PDS Archive,
along with a revised version of the classification by
\citet{2010A&A...510A..43C}, which includes the B class. 
The classification can be used to perceive how any observation compares
with the template spectra of a given class in terms of band depth and
spectral slope, allowing more refined searches
inside the large data sets as the SDSSMOC. For this purpose, all SDSSMOC
observations were further classified as \emph{Red}, \emph{Blue} or
\emph{Red+Blue}. 
In this sense, the scheme here described, although carrying less information, 
is more appropriate for observations outside the formal
taxonomic boundaries.


In conclusion, we were able to distinguish the basic asteroid spectra trends by analyzing a
selected sample of high-quality observations from SDSSMOC data. 
The addition of taxonomic locus helped us to give a compositional interpretation to both trends. 
Although the scheme was derived from a specific dataset, any observation in a similar photometric system, 
or any low-resolution spectrum convoluted with SDSS
bandpasses, may also be classified.

\begin{acknowledgements}
The authors thanks CNPq, process no.402085/2012-4, for the support.
FAPERJ and CAPES are also acknowledged for diverse grants and fellowships 
to D.L., J.M.C., and P.H.H.
\end{acknowledgements}

\bibliographystyle{aa}
\bibliography{sdssmoc}

\begin{thebibliography}{47}
\expandafter\ifx\csname natexlab\endcsname\relax\def\natexlab#1{#1}\fi

\bibitem[{{Barucci} {et~al.}(1987){Barucci}, {Capria}, {Coradini}, \&
  {Fulchignoni}}]{1987Icar...72..304B}
{Barucci}, M.~A., {Capria}, M.~T., {Coradini}, A., \& {Fulchignoni}, M. 1987,
  Icarus, 72, 304

\bibitem[{{Bell}(1986)}]{1986LPI....17..985B}
{Bell}, J.~F. 1986, in Lunar and Planetary Science Conference, Vol.~17, Lunar
  and Planetary Science Conference, 985--986

\bibitem[{{Bentley}(1979)}]{1979IEEE...4B}
{Bentley}, J.~L. 1979, IEEE Transactions on Software Engeneering, 4

\bibitem[{{Bowell} {et~al.}(1978){Bowell}, {Chapman}, {Gradie}, {Morrison}, \&
  {Zellner}}]{1978Icar...35..313B}
{Bowell}, E., {Chapman}, C.~R., {Gradie}, J.~C., {Morrison}, D., \& {Zellner},
  B. 1978, Icarus, 35, 313

\bibitem[{{Bowell} {et~al.}(1994){Bowell}, {Muinonen}, \&
  {Wasserman}}]{1994IAUS..160..477B}
{Bowell}, E., {Muinonen}, K., \& {Wasserman}, L.~H. 1994, in IAU Symposium,
  Vol. 160, Asteroids, Comets, Meteors 1993, ed. {A.~Milani, M.~di Martino, \&
  A.~Cellino}, 477--+

\bibitem[{{Bus} \& {Binzel}(2002)}]{2002Icar..158..146B}
{Bus}, S.~J. \& {Binzel}, R.~P. 2002, Icarus, 158, 146

\bibitem[{{Carvano} {et~al.}(2010){Carvano}, {Hasselmann}, {Lazzaro}, \&
  {Moth{\'e}-Diniz}}]{2010A&A...510A..43C}
{Carvano}, J.~M., {Hasselmann}, P.~H., {Lazzaro}, D., \& {Moth{\'e}-Diniz}, T.
  2010, Astronomy and Astrophysics, 510, A43+

\bibitem[{{Carvano} {et~al.}(2003){Carvano}, {Moth{\'e}-Diniz}, \&
  {Lazzaro}}]{2003Icar..161..356C}
{Carvano}, J.~M., {Moth{\'e}-Diniz}, T., \& {Lazzaro}, D. 2003, \icarus, 161,
  356

\bibitem[{{Chapman} {et~al.}(1975){Chapman}, {Morrison}, \&
  {Zellner}}]{1975Icar...25..104C}
{Chapman}, C.~R., {Morrison}, D., \& {Zellner}, B. 1975, Icarus, 25, 104

\bibitem[{{Delbo'} {et~al.}(2012){Delbo'}, {Gayon-Markt}, {Busso}, {Brown},
  {Galluccio}, {Ordenovic}, {Bendjoya}, \& {Tanga}}]{2012P&SS...73...86D}
{Delbo'}, M., {Gayon-Markt}, J., {Busso}, G., {et~al.} 2012, \planss, 73, 86

\bibitem[{{DeMeo} \& {Binzel}(2008)}]{2008Icar..194..436D}
{DeMeo}, F. \& {Binzel}, R.~P. 2008, \icarus, 194, 436

\bibitem[{{DeMeo} {et~al.}(2009){DeMeo}, {Binzel}, {Slivan}, \&
  {Bus}}]{2009Icar..202..160D}
{DeMeo}, F.~E., {Binzel}, R.~P., {Slivan}, S.~M., \& {Bus}, S.~J. 2009, Icarus,
  202, 160

\bibitem[{{Fukugita} {et~al.}(1996){Fukugita}, {Ichikawa}, {Gunn}, {Doi},
  {Shimasaku}, \& {Schneider}}]{1996AJ....111.1748F}
{Fukugita}, M., {Ichikawa}, T., {Gunn}, J.~E., {et~al.} 1996, Astrophisical
  Journal, 111, 1748

\bibitem[{{Gaffey}(2010)}]{2010Icar..209..564G}
{Gaffey}, M.~J. 2010, Icarus, 209, 564

\bibitem[{{Gunn} {et~al.}(1998){Gunn}, {Carr}, {Rockosi}, {Sekiguchi}, {Berry},
  {Elms}, {de Haas}, {Ivezi{\'c}}, {Knapp}, {Lupton}, {Pauls}, {Simcoe},
  {Hirsch}, {Sanford}, {Wang}, {York}, {Harris}, {Annis}, {Bartozek},
  {Boroski}, {Bakken}, {Haldeman}, {Kent}, {Holm}, {Holmgren}, {Petravick},
  {Prosapio}, {Rechenmacher}, {Doi}, {Fukugita}, {Shimasaku}, {Okada}, {Hull},
  {Siegmund}, {Mannery}, {Blouke}, {Heidtman}, {Schneider}, {Lucinio}, \&
  {Brinkman}}]{1998AJ....116.3040G}
{Gunn}, J.~E., {Carr}, M., {Rockosi}, C., {et~al.} 1998, \aj, 116, 3040

\bibitem[{{Hasselmann} {et~al.}(2011){Hasselmann}, {Carvano}, \&
  {Lazzaro}}]{2011PDSS..145.....H}
{Hasselmann}, P.~H., {Carvano}, J.~M., \& {Lazzaro}, D. 2011, NASA Planetary
  Data System, 145

\bibitem[{{Hasselmann} {et~al.}(2013){Hasselmann}, {Carvano}, \&
  {Lazzaro}}]{2013scpy.conf...48H}
{Hasselmann}, P.~H., {Carvano}, J.~M., \& {Lazzaro}, D. 2013, in Proceedings of
  the 12th Python in Science Conference (SciPy 2013), p. 48-55, 48--55

\bibitem[{{Ivezi{\'c}} {et~al.}(2002){Ivezi{\'c}}, {Juric}, {Lupton},
  {Tabachnik}, \& {Quinn}}]{2002SPIE.4836...98I}
{Ivezi{\'c}}, Z., {Juric}, M., {Lupton}, R.~H., {Tabachnik}, S., \& {Quinn}, T.
  2002, in Society of Photo-Optical Instrumentation Engineers (SPIE) Conference
  Series, Vol. 4836, Society of Photo-Optical Instrumentation Engineers (SPIE)
  Conference Series, ed. J.~A. {Tyson} \& S.~{Wolff}, 98--103

\bibitem[{{Ivezic} {et~al.}(2010){Ivezic}, {Juric}, {Lupton}, {Tabachnik},
  {Quinn}, \& {The SDSS Collaboration}}]{2010PDSS..124.....I}
{Ivezic}, Z., {Juric}, M., {Lupton}, R.~H., {et~al.} 2010, NASA Planetary Data
  System, 124

\bibitem[{Jolliffe(1986)}]{PCAJolliffe}
Jolliffe, I.~T. 1986, Principal Component Analysis (New York: Springer-Verlag)

\bibitem[{{Jones} {et~al.}(2014){Jones}, {Ivezic}, {Malhotra}, {Becker},
  {Fernandez}, {Myers}, {Solontoi}, \& {Parker}}]{2014DPS....4621413J}
{Jones}, R.~L., {Ivezic}, Z., {Malhotra}, R., {et~al.} 2014, in AAS/Division
  for Planetary Sciences Meeting Abstracts, Vol.~46, AAS/Division for Planetary
  Sciences Meeting Abstracts, 214.13

\bibitem[{{Lantz} {et~al.}(2013){Lantz}, {Clark}, {Barucci}, \&
  {Lauretta}}]{2013A&A...554A.138L}
{Lantz}, C., {Clark}, B.~E., {Barucci}, M.~A., \& {Lauretta}, D.~S. 2013, \aap,
  554, A138

\bibitem[{{Lazzaro} {et~al.}(2004){Lazzaro}, {Angeli}, {Carvano},
  {Moth{\'e}-Diniz}, {Duffard}, \& {Florczak}}]{2004Icar..172..179L}
{Lazzaro}, D., {Angeli}, C.~A., {Carvano}, J.~M., {et~al.} 2004, Icarus, 172,
  179

\bibitem[{{Lupton} {et~al.}(1999){Lupton}, {Gunn}, \&
  {Szalay}}]{1999AJ....118.1406L}
{Lupton}, R.~H., {Gunn}, J.~E., \& {Szalay}, A.~S. 1999, Astrophysical Journal,
  118, 1406

\bibitem[{{Miller} {et~al.}(2014){Miller}, {De Ruette}, {Harlow}, {Domingue},
  \& {Savin}}]{2014AAS...22411911M}
{Miller}, K.~A., {De Ruette}, N., {Harlow}, G., {Domingue}, D.~L., \& {Savin},
  D.~W. 2014, in American Astronomical Society Meeting Abstracts, Vol. 224,
  American Astronomical Society Meeting Abstracts 224, 119.11

\bibitem[{{Moroz} {et~al.}(1996){Moroz}, {Fisenko}, {Semjonova}, {Pieters}, \&
  {Korotaeva}}]{1996Icar..122..366M}
{Moroz}, L.~V., {Fisenko}, A.~V., {Semjonova}, L.~F., {Pieters}, C.~M., \&
  {Korotaeva}, N.~N. 1996, Icarus, 122, 366

\bibitem[{{Morrison} \& {Zellner}(2007)}]{TRIADRAD}
{Morrison}, D. \& {Zellner}, B. 2007, NASA Planetary Data System

\bibitem[{{Neese}(2010)}]{2010PDSS..123.....N}
{Neese}, C. 2010, NASA Planetary Data System, 123

\bibitem[{{Nesvorn{\'y}} {et~al.}(2005){Nesvorn{\'y}}, {Jedicke}, {Whiteley},
  \& \v{Z}. {Ivezi{\'c}}}]{2005Icar..173..132N}
{Nesvorn{\'y}}, D., {Jedicke}, R., {Whiteley}, R.~J., \& \v{Z}. {Ivezi{\'c}}.
  2005, Icarus, 173, 132

\bibitem[{{Parker} {et~al.}(2008){Parker}, \v{Z}. {Ivezi{\'c}}, {Juri{\'c}},
  {Lupton}, {Sekora}, \& {Kowalski}}]{2008Icar..198..138P}
{Parker}, A., \v{Z}. {Ivezi{\'c}}, {Juri{\'c}}, M., {et~al.} 2008, Icarus, 198,
  138

\bibitem[{Pedregosa {et~al.}(2011)Pedregosa, Varoquaux, Gramfort, Michel,
  Thirion, Grisel, Blondel, Prettenhofer, Weiss, Dubourg, Vanderplas, Passos,
  Cournapeau, Brucher, Perrot, \& Duchesnay}]{scikit-learn}
Pedregosa, F., Varoquaux, G., Gramfort, A., {et~al.} 2011, Journal of Machine
  Learning Research, 12, 2825

\bibitem[{{Perna} {et~al.}(2014){Perna}, {Alvarez-Candal}, {Fornasier}, {Ka{\v
  n}uchov{\'a}}, {Giuliatti Winter}, {Vieira Neto}, \&
  {Winter}}]{2014A&A...568L...6P}
{Perna}, D., {Alvarez-Candal}, A., {Fornasier}, S., {et~al.} 2014, \aap, 568,
  L6

\bibitem[{Press {et~al.}(1988)Press, Flannery, Teukolsky, \&
  Vetterling}]{Press:1988:NRC:42249}
Press, W.~H., Flannery, B.~P., Teukolsky, S.~A., \& Vetterling, W.~T. 1988,
  Numerical Recipes in C: The Art of Scientific Computing (New York, NY, USA:
  Cambridge University Press)

\bibitem[{{Reddy} {et~al.}(2011){Reddy}, {Carvano}, {Lazzaro}, {Michtchenko},
  {Gaffey}, {Kelley}, {Moth{\'e}-Diniz}, {Alvarez-Candal}, {Moskovitz},
  {Cloutis}, \& {Ryan}}]{2011Icar..216..184R}
{Reddy}, V., {Carvano}, J.~M., {Lazzaro}, D., {et~al.} 2011, \icarus, 216, 184

\bibitem[{{Reddy} {et~al.}(2014){Reddy}, {Sanchez}, {Bottke}, {Cloutis},
  {Izawa}, {O'Brien}, {Mann}, {Cuddy}, {Le Corre}, {Gaffey}, \&
  {Fujihara}}]{2014Icar..237..116R}
{Reddy}, V., {Sanchez}, J.~A., {Bottke}, W.~F., {et~al.} 2014, \icarus, 237,
  116

\bibitem[{{Roig} \& {Gil-Hutton}(2006)}]{2006Icar..183..411R}
{Roig}, F. \& {Gil-Hutton}, R. 2006, Icarus, 183, 411

\bibitem[{{Sanchez} {et~al.}(2012){Sanchez}, {Reddy}, {Nathues}, {Cloutis},
  {Mann}, \& {Hiesinger}}]{2012Icar..220...36S}
{Sanchez}, J.~A., {Reddy}, V., {Nathues}, A., {et~al.} 2012, Icarus, 220, 36

\bibitem[{{Solontoi} {et~al.}(2010){Solontoi}, \v{Z}. {Ivezi{\'c}}, {West},
  {Claire}, {Juri{\'c}}, {Becker}, {Jones}, {Hall}, {Kent}, {Lupton}, {Knapp},
  {Quinn}, {Gunn}, {Schneider}, \& {Loomis}}]{2010Icar..205..605S}
{Solontoi}, M., \v{Z}. {Ivezi{\'c}}, {West}, A.~A., {et~al.} 2010, Icarus, 205,
  605

\bibitem[{{Stoughton} {et~al.}(2002){Stoughton}, {Lupton}, {Bernardi},
  {Blanton}, {Burles}, {Castander}, {Connolly}, {Eisenstein}, {Frieman},
  {Hennessy}, {Hindsley}, \v{Z}. {Ivezi{\'c}}, {Kent}, {Kunszt}, {Lee},
  {Meiksin}, {Munn}, {Newberg}, {Nichol}, {Nicinski}, {Pier}, {Richards},
  {Richmond}, {Schlegel}, {Smith}, {Strauss}, {SubbaRao}, {Szalay}, {Thakar},
  {Tucker}, {Vanden Berk}, {Yanny}, {Adelman}, {Anderson}, {Anderson}, {Annis},
  {Bahcall}, {Bakken}, {Bartelmann}, {Bastian}, {Bauer}, {Berman},
  {B{\"o}hringer}, {Boroski}, {Bracker}, {Briegel}, {Briggs}, {Brinkmann},
  {Brunner}, {Carey}, {Carr}, {Chen}, {Christian}, {Colestock}, {Crocker},
  {Csabai}, {Czarapata}, {Dalcanton}, {Davidsen}, {Davis}, {Dehnen},
  {Dodelson}, {Doi}, {Dombeck}, {Donahue}, {Ellman}, {Elms}, {Evans}, {Eyer},
  {Fan}, {Federwitz}, {Friedman}, {Fukugita}, {Gal}, {Gillespie}, {Glazebrook},
  {Gray}, {Grebel}, {Greenawalt}, {Greene}, {Gunn}, {de Haas}, {Haiman},
  {Haldeman}, {Hall}, {Hamabe}, {Hansen}, {Harris}, {Harris}, {Harvanek},
  {Hawley}, {Hayes}, {Heckman}, {Helmi}, {Henden}, {Hogan}, {Hogg}, {Holmgren},
  {Holtzman}, {Huang}, {Hull}, {Ichikawa}, {Ichikawa}, {Johnston}, {Kauffmann},
  {Kim}, {Kimball}, {Kinney}, {Klaene}, {Kleinman}, {Klypin}, {Knapp},
  {Korienek}, {Krolik}, {Kron}, {Krzesi{\'n}ski}, {Lamb}, {Leger},
  {Limmongkol}, {Lindenmeyer}, {Long}, {Loomis}, {Loveday}, {MacKinnon},
  {Mannery}, {Mantsch}, {Margon}, {McGehee}, {McKay}, {McLean}, {Menou},
  {Merelli}, {Mo}, {Monet}, {Nakamura}, {Narayanan}, {Nash}, {Neilsen},
  {Newman}, {Nitta}, {Odenkirchen}, {Okada}, {Okamura}, {Ostriker}, {Owen},
  {Pauls}, {Peoples}, {Peterson}, {Petravick}, {Pope}, {Pordes}, {Postman},
  {Prosapio}, {Quinn}, {Rechenmacher}, {Rivetta}, {Rix}, {Rockosi}, {Rosner},
  {Ruthmansdorfer}, {Sandford}, {Schneider}, {Scranton}, {Sekiguchi}, {Sergey},
  {Sheth}, {Shimasaku}, {Smee}, {Snedden}, {Stebbins}, {Stubbs}, {Szapudi},
  {Szkody}, {Szokoly}, {Tabachnik}, {Tsvetanov}, {Uomoto}, {Vogeley}, {Voges},
  {Waddell}, {Walterbos}, i.~{Wang}, {Watanabe}, {Weinberg}, {White}, {White},
  {Wilhite}, {Wolfe}, {Yasuda}, {York}, {Zehavi}, \&
  {Zheng}}]{2002AJ....123..485S}
{Stoughton}, C., {Lupton}, R.~H., {Bernardi}, M., {et~al.} 2002, Astrophisical
  Journal, 123, 485

\bibitem[{{Tedesco} {et~al.}(1989){Tedesco}, {Williams}, {Matson}, {Weeder},
  {Gradie}, \& {Lebofsky}}]{1989AJ.....97..580T}
{Tedesco}, E.~F., {Williams}, J.~G., {Matson}, D.~L., {et~al.} 1989,
  Astrophysical Journal, 97, 580

\bibitem[{{Tholen}(1984)}]{1984PhDT.........3T}
{Tholen}, D.~J. 1984, PhD thesis, Arizona Univ., Tucson.

\bibitem[{{Tholen} \& {Barucci}(1989)}]{1989aste.conf..298T}
{Tholen}, D.~J. \& {Barucci}, M.~A. 1989, in Asteroids II, ed. {R.~P.~Binzel,
  T.~Gehrels, \& M.~S.~Matthews}, 298--315

\bibitem[{\v{Z}. {Ivezi{\'c}} {et~al.}(2001)\v{Z}. {Ivezi{\'c}}, {Tabachnik},
  {Rafikov}, {Lupton}, {Quinn}, {Hammergren}, {Eyer}, {Chu}, {Armstrong},
  {Fan}, {Finlator}, {Geballe}, {Gunn}, {Hennessy}, {Knapp}, {Leggett}, {Munn},
  {Pier}, {Rockosi}, {Schneider}, {Strauss}, {Yanny}, {Brinkmann}, {Csabai},
  {Hindsley}, {Kent}, {Lamb}, {Margon}, {McKay}, {Smith}, {Waddel}, {York}, \&
  {the SDSS Collaboration}}]{2001AJ....122.2749I}
\v{Z}. {Ivezi{\'c}}, {Tabachnik}, S., {Rafikov}, R., {et~al.} 2001,
  Astrophysical Journal, 122, 2749

\bibitem[{{Wall} {et~al.}(2002){Wall}, {Rechtsteiner}, \&
  {Rocha}}]{2002physics...8101W}
{Wall}, M.~E., {Rechtsteiner}, A., \& {Rocha}, L.~M. 2002, ArXiv Physics
  e-prints

\bibitem[{{Warell} \& {Lagerkvist}(2007)}]{2007A&A...467..749W}
{Warell}, J. \& {Lagerkvist}, C.-I. 2007, Astronomy and Astrophysics, 467, 749

\bibitem[{{Yang} \& {Jewitt}(2010)}]{2010AJ....140..692Y}
{Yang}, B. \& {Jewitt}, D. 2010, \aj, 140, 692

\bibitem[{{Zellner} {et~al.}(1985){Zellner}, {Tholen}, \&
  {Tedesco}}]{1985Icar...61..355Z}
{Zellner}, B., {Tholen}, D.~J., \& {Tedesco}, E.~F. 1985, Icarus, 61, 355

\end{thebibliography}

\end{document}